\documentclass[reprint, floatfix]{revtex4-1}

\usepackage{graphicx}
\usepackage{amssymb}
\usepackage{amsmath}
\usepackage{epsfig}
\usepackage{chemarr}
\usepackage{url}
\usepackage{natbib}

\usepackage{dcolumn}
\usepackage{bm}
\usepackage{color}

\usepackage{comment}
\usepackage[]{changes}
\definechangesauthor[name={Arvind's remarks}, color=red]{Arvind}
\definechangesauthor[name={Nachi's remarks}, color=orange]{Nachi}

\begin{document}

\title{Learned multi-stability in mechanical networks}

\author{Menachem Stern, Matthew B. Pinson, Arvind Murugan}

\affiliation{Physics Department and the James Franck Institute,
	University of Chicago, Chicago, IL 60637}

\begin{abstract}
We contrast the distinct frameworks of materials design and physical learning in creating elastic networks with desired stable states.
In design, the desired states are specified in advance and material parameters can be optimized on a computer with this knowledge. In learning, the material physically experiences the desired stable states in sequence, changing the material so as to stabilize each additional state. We show that while designed states are stable in networks of linear Hookean springs, sequential learning requires specific non-linear elasticity. We find that such non-linearity stabilizes states in which strain is zero in some springs and large in others, thus playing the role of Bayesian priors used in sparse statistical regression. Our model shows how specific material properties allow continuous learning of new functions through deployment of the material itself.
\end{abstract}
\maketitle

Materials design is generally predicated on knowing the desired material behavior at the time of design. If an adaptable material with multiple behaviors is desired, all potential desired behaviors are usually specified in advance. As a result, we can optimize design parameters compatible with all of the specified desired behaviors. Among mechanical metamaterials, such design has been fruitfully used to create materials that switch from being soft to stiff, transparent to opaque or energy absorbing to elastic, by simply switching between different stable geometric states of the material \cite{silverberg2015origami, Waitukaitis2015-rw, overvelde2016three, shan2015multistable, bertoldi2017flexible, steinbach2016bistable,Wu2018-ad, Yang2018-kv, Che2017-bu, Yang2018-ep, Fu2018-ya}.

Here, we explore an alternative approach, where a material \emph{learns} desired behaviors on the fly by physically experiencing such behaviors in sequence, e.g., by being held in each desired state for a period of time. Such a learning framework for materials offers many complementary strengths to the conventional design framework. For example, the precise behaviors needed can be inferred from the actual conditions of use in real time, instead of an \textit{a priori} specification. New functionalities can be gained during, and due to, use.
Such benefits have made learning a powerful framework in neuroscience and artificial neural networks, but this framework is relatively unexplored in the context of materials \cite{rocks2017designing, rocks2018limits, hexner2018role}.

However, learning in the context of materials presents challenges in addition to such opportunities. In the learning framework, the desired behaviors are not all known ahead of time but presented sequentially. Thus material parameters to encode each desired behavior must be chosen independently without knowledge of future desired behaviors. Most critically, each stored behavior or state needs to survive the parameter changes due to the subsequent learned behaviors and not be overwritten by them. It is not clear what kinds of material properties and interactions would allow such sequential learning of multiple behaviors.

In this work we contrast the requirements for design and learning of multiple stable states in a simple elastic network. In the design model, we search over all spring constants on a computer to stabilize a set of states that are specified beforehand. In the learning model, the desired states are learned in sequence by example, placing the material in each of these states for a period of time. During this time, stabilizing elastic rods or springs with a rest length grow between particles within some distance in space, {mimicking the seeded growth of microtubules \cite{Hess2017-gi} or self-assembling DNA nanotubes \cite{mohammed2013directing}}. Thus, in contrast to design, the learning model is constrained by locality in space and time -- material parameters are modified only by the local geometry of the current configuration being experienced \cite{rocks2018limits, hexner2018role}.

As a direct consequence, we find that successful learning requires non-linear elasticity of a specific type. Parameterizing the elastic energy of springs in the network as $E \sim x^\xi$ for large extensions $x$, we find that our design procedure is optimal for $\xi \approx 2$ (Hooke's law) but learning requires $0 < \xi \leq 1$. Such nonlinear springs have been demonstrated using metamaterial designs \cite{IsobeNonlinearSpring,ChenNonlinearSpring}. We relate this distinction to the way springs are unequally strained in a learned state -- springs learned for that state are nearly unstrained while all other springs are highly strained. Such `sparse' strain profiles are stabilized by $\xi \leq 1$ springs but not for $\xi >1$.

We establish these results by relating spring non-linearity to Bayesian priors used in statistical regression; such priors can pick out sparse solutions to equations in which some variables are exactly zero. Much in the way Bayesian priors dictate sparsity in statistical regression, the non-linearity of springs dictates that information about each learned state is localized in the material. We hope our analysis of a simple mechanical model will stimulate further work on the conditions under which materials can learn new functionalities on the fly.

\begin{figure*}	
\includegraphics[width=1.0\linewidth]{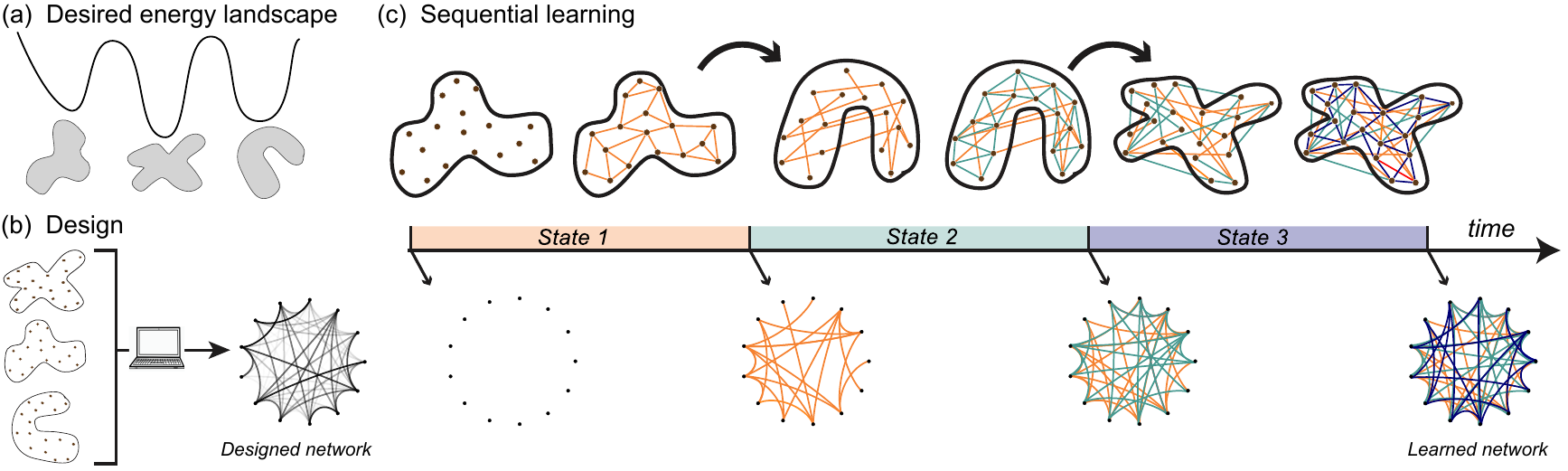}
\caption{Designing vs learning multiple states: (a) We seek to create an elastic network with specific stable configurations. (b) In the design approach, all desired states are specified beforehand and then network parameters (connectivity, spring constants, rest lengths) are optimized to stabilize these states. (c) In learning, the material is physically placed in the desired states in sequence and the network grows incrementally according to the local geometry of that state (Eq.~\ref{eq:LearningRule}). Hence information about each desired state is localized to only a fraction of network links (different colors). For learning to succeed, network changes due to learning state $2$ should not interfere with the stability of state $1$ or vice-versa. \label{fig:schematic}}
\end{figure*}

\section*{Results}

We seek to create an elastic network of springs connecting $N$ particles in $2$ dimensions, such that the network has $M$ desired stable states (Fig.~\ref{fig:schematic}a). Each desired stable state $m=1,\ldots M$ is specified by the positions $\textbf{x}^{(m)}$ of the $N$ particles (up to rigid body translations and rotations).

In our design model, we connect the $N$ particles by Hookean (linear) springs, and solve an optimization problem for spring constants $k_{ij}$ and rest lengths $l_{ij}$ that minimizes residual forces at each of the desired configurations $\textbf{x}^{(m)}$ (Fig.~\ref{fig:schematic}b); see Supplementary Note 1 for details.

In the learning model, desired stable states are acquired by sequentially placing the material in the desired configurations (Fig.~\ref{fig:schematic}c). When left in a configuration $\textbf{x}^{(1)}$ for a length of time, unstretched elastic rods grow between every pair of particles $i,j$ at a rate $f(r_{ij})$ set by their separation $r_{ij}$; we assume that $f$ vanishes rapidly outside of a characteristic length scale $R$, so only nodes within a distance less than $R$ are stabilized by such rods. {Such elastic elements that grow between specific sites are found both in living systems  (e.g., microtubules growing between centrosomes and centromeres \cite{Dogterom2013-vk,Hess2017-gi}) and in synthetic systems (e.g., self-assembling DNA nanotubes \cite{mohammed2013directing} growing between seeds)}. 

Since the number of rods grows with time, the effective spring constant for the set of rods connecting two particles $i,j$ grows with time and is given by,
\begin{equation}
    \frac{d k^{\mbox{eff}}_{ij}}{dt} = k_0 f(r_{ij}).   
\label{eq:LearningRule}
\end{equation}

Here $k_0$ is the spring constant of each rod, whose rest length $l_{ij}$ is assumed equal to the particle separation $r_{ij}$, i.e., rods are unstretched in the desired state. In simulations, we pick $f$ to be a step function of range $R$, $f(r< R) =1, f(r>R)=0$.

Equation~\ref{eq:LearningRule} describes the learning rule for this material; the effective spring constant and rest length between two particles $i,j$ is determined by the geometric configurations experienced by the material and the amount of time spent in each configuration. When the material is deformed and held in a second distinct configuration $\textbf{x}^{(2)}$, additional rods start growing between the particles according to their positions in the new configuration. In some cases, two particles can be joined by multiple springs with different rest lengths.

\subsection*{Linear and non-linear elasticity}

\begin{figure}	
\includegraphics[width=1.0\linewidth]{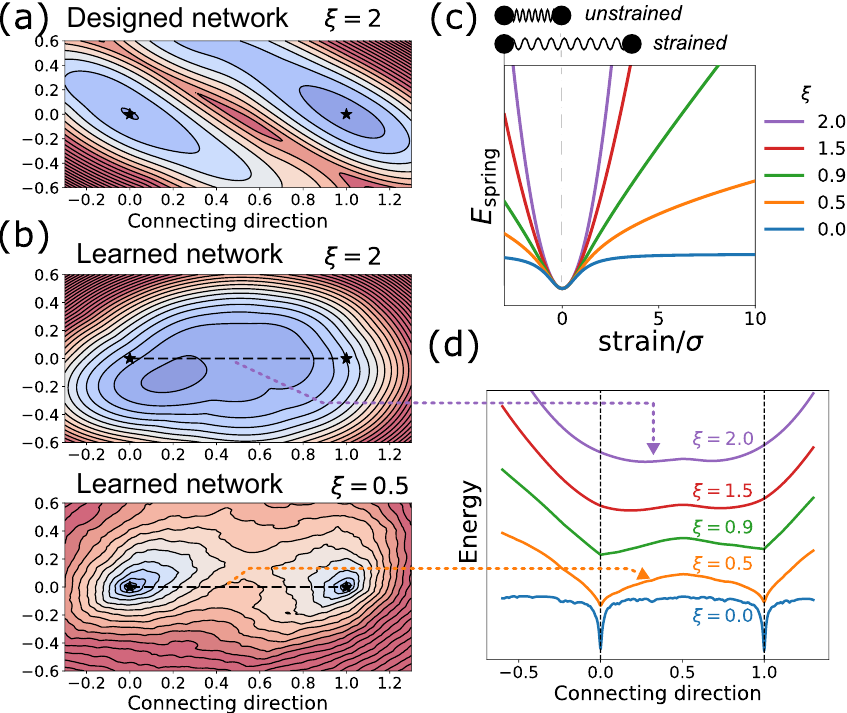}
\caption{Non-linear interactions are essential for learning multiple states in sequence. (a) Energy landscape of a designed network with linear ($\xi=2$) springs successfully stabilizes desired states (black stars). (b) In the learned network, linear ($\xi=2$) springs learned for each desired state destabilize the other state, but non-linear ($\xi=0.5$) learned springs stabilize both desired states. (c,d) Repeating learning for non-linear springs with $E \sim s^\xi$, we find that learned states overwrite each other for $\xi> 1$ but are protected from each other with sufficiently non-linear $\xi \le 1$ springs.
\label{fig:nonlinear}}
\end{figure}

We ran the design and learning algorithms using rods with linear Hookean elasticity, i.e., with elastic energy $E_{ij} \sim k_0 s_{ij}^2$ when strained by $s_{ij}$.  The design algorithm, when run on a pair of randomly generated desired states $\textbf{x}^{(1)}$ and $\textbf{x}^{(2)}$ of 10 particles, resulted in an elastic network with two stable minima that resemble $\textbf{x}^{(1)}$ and $\textbf{x}^{(2)}$, as seen Fig.~\ref{fig:nonlinear}a. 
These states can be retrieved by any initial condition within wide attractor regions around $\textbf{x}^{(1)}, \textbf{x}^{(2)}$.

In contrast, learning the same two states $\textbf{x}^{(1)},\textbf{x}^{(2)}$ with linear springs fails (Fig.~\ref{fig:nonlinear}b); the two desired states are not stable minima of the learned network. The rods grown to encode state $\textbf{x}^{(1)}$ destabilize, or overwrite, state $\textbf{x}^{(2)}$ and vice-versa. Initial conditions near either $\textbf{x}^{(1)}$ or $\textbf{x}^{(2)}$ relax to new minima very different from  $\textbf{x}^{(1)}$, $\textbf{x}^{(2)}$.

Why do linear springs allow stabilization of multiple states with design but not with sequential learning? In design, the desired configurations are known ahead of time and so each spring's parameters can be chosen cognizant of all desired configurations. In fact, one can check that changing one of the desired states, e.g., $\textbf{x}^{(1)} \to \textbf{x}^{(1)} + \delta \textbf{x}^{(1)}$ changes stiffness $k_{ij}$ and rest length $l_{ij}$ for \emph{all} springs. In this sense, information about each desired state is stored in every spring. 

However, in a learning model capable of acquiring arbitrary stable states in sequence, the parameter changes made to store a state  $\textbf{x}^{(m)}$ cannot depend on the details of future desired states \cite{Hebbian1}, and indeed, in this model, does not depend on past encoded states either. That is, changing a desired configuration, e.g., $\textbf{x}^{(m)} \to \textbf{x}^{(m)} + \delta \textbf{x}^{(m)}$ changes the spring parameters $k_{ij}$, $l_{ij}$ only for springs grown while learning state $m$. Thus, information about each stored state is confined to a subset of springs. 

Consequently, to stabilize a state $\textbf{x}^{(m)}$, the elastic dynamics should only attempt to minimize strain to zero in a subset of all springs while leaving all other springs stretched arbitrarily as needed. However, the mechanics cannot possibly know which subset of springs was learned to stabilize a particular state $\textbf{x}^{(m)}$ and thus which subset to satisfy.

A clue to solving this problem comes from sparse regression \cite{Lasso,zou2005regularization}. 
As an example, consider an under-determined problem $A \mathbf{s} = b$ for a vector $\mathbf{s}$. If we know \emph{a priori} that an $\mathbf{s}$ exists which has some components that are strictly zero and others non-zero, we can find such `sparse' solutions $\mathbf{s}$ by adding a `Bayesian prior' $||\mathbf{s}||^\xi = \sum_i s_i^\xi$ to the least squares loss function, 
\begin{equation}
    E = ||A \mathbf{s} - b||^2 + ||\mathbf{s}||^\xi 
\label{eq:LASSO}
\end{equation}

\begin{figure}	
\includegraphics[width=1.0\linewidth]{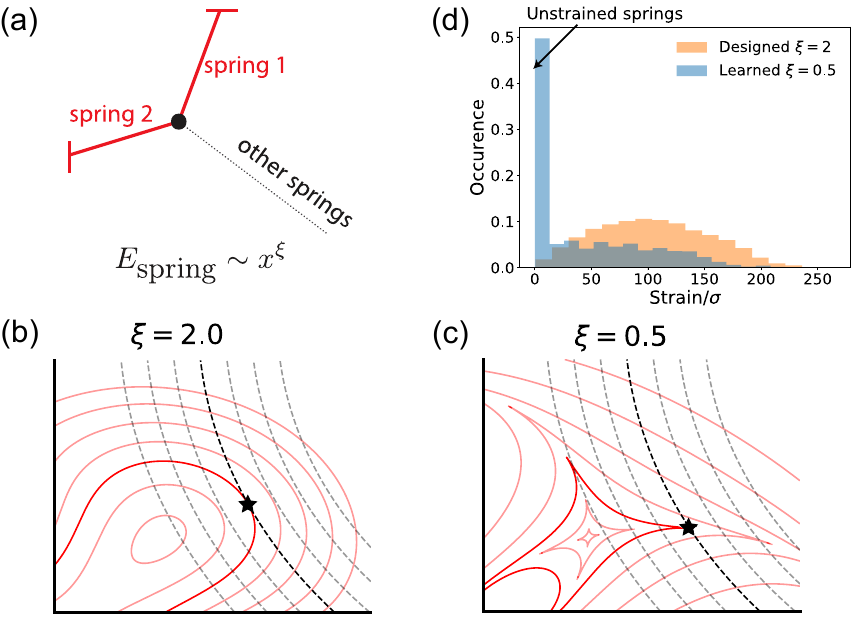}
\caption{Non-linear springs apply a Bayesian prior to the strain distribution. (a) The  energy of the two red springs is represented by red contours, that of all other springs by black contours. The whole system's energy minima will be at points where these contours are tangent to one another.  (b) If $\xi > 1$, the minimum is at a generic point with no special features. (c) If $\xi \leq 1$, the minimum is very likely to be at a red cusp, corresponding to a configuration in which one of the red springs is unstrained. (d) Typical stable states of a large $N = 100$ network have many unstrained springs if and only if $\xi \leq 1$.}
\label{fig:sparse}
\end{figure}

and then minimizing $E$ \cite{Lasso,SparseRegression}. If $\xi \leq 1$, such a Bayesian prior $||\mathbf{s}||^\xi $ biases the search towards solutions $\mathbf{s}$ in which some elements of $\mathbf{s}$ are strictly zero while others are non-zero (i.e., `sparse' solutions). We emphasize that the Bayesian prior $s^\xi$ contains no information about which components of $s$ are to be set to zero; rather, it biases regression towards such solutions and away from generic solutions in which all entries of $\mathbf{s}$ are non-zero.

We employ a similar strategy here by identifying $\mathbf{s}$ above with the vector of strains in different springs. Let us assume that the network spring energies take a non-linear form, 

\begin{equation}
  E(s) \sim k_0 \frac{s^2}{(\sigma^2+s^2)^{1-0.5\xi}},
  \label{eq:Energy}
\end{equation}    

where $k_0$ is the spring constant and $s_{ij} \equiv (r_{ij} - l_{ij})$ is the strain relative to rest length $l_{ij}$.  $\xi$ parameterizes the non-linearity (Fig.~\ref{fig:nonlinear}c); $\xi=2$ is a linear Hookean spring while $\xi < 2$ springs have softer restoring forces at large distances, $E \sim s^\xi$. 
Finally, $\sigma$ is a small length scale within which the interaction is linear for any $\xi$ and is introduced to keep the model realistic, reflecting  practical realizations of non-linear $\xi < 2$ springs  \cite{IsobeNonlinearSpring,ChenNonlinearSpring}; our results below hold for $\sigma \to 0$ as well.  See Supplementary Note 2 for details. 

We repeated the same learning procedure on the same states as earlier - but with non-linear springs $\xi< 2$.  While the results for $1 < \xi <2$ are qualitatively similar to linear springs $\xi = 2$, $\xi < 1$ shows qualitatively different results -- learning succeeds in stabilizing multiple states (Fig.~\ref{fig:nonlinear}b,d).

How do we understand this result? It is clear that forces due to $\xi<1$ springs diminish with strain and thus weaken the effect of strained springs that code for other states. However, the analogy with Bayesian priors goes further by explaining the sharp change in behavior at $\xi = 1$ due to the non-analytic nature of $s^\xi$. Following work in sparse regression \cite{Lasso}, in Fig.~\ref{fig:sparse}b,c, we plot the energy contours for the red springs shown, where the two red springs have incompatible rest lengths. 
The constant energy contours are cusped for $\xi < 1$ but not $\xi > 1$. At the cusps, one of the two red springs is completely unstrained while the other contains all the strain. When minimized in conjunction with other springs (dashed black contours), minima are exceedingly likely to be at cusps for $\xi < 1$, where strain is localized to one spring.

Thus, non-linear $\xi < 1$ springs stabilize states with bimodal strain distributions - some springs are highly strained while others are unstrained. To complete the analogy with sparse regression,  note that the energy of the system in Fig.~\ref{fig:sparse} resembles Eq.~\ref{eq:LASSO}. Let $\mathbf{F}^{ext}$ represent forces on the particle in Fig.~\ref{fig:nonlinear}a due to the black springs (assumed constant for simplicity). 
In the limit of small core sizes $\sigma\rightarrow 0$, the red spring energies are given by $E(r)=k (r-l)^\xi\equiv k s^\xi$, so that the total energy of the subsystem shown is,
\begin{equation}
    E = -\mathbf{F}^{ext}\cdot \mathbf{x} + k\sum_{\mbox{red}} s_a^\xi = -\mathbf{F}^{ext}\cdot \mathbf{x} + k \vert\vert \mathbf{s} \vert\vert^\xi,
\label{eq:Regularization}
\end{equation}

where $\vert\vert \mathbf{s} \vert\vert^\xi$ is the $\xi$-norm of the strain vector $\mathbf{s}$ for the red springs. The non-linear elastic energy has the analytic form of sparse regression, Eq.~\ref{eq:LASSO}, and thus one of the red springs is unstrained in each stable minimum.  
Note that the springs now play a dual role, both providing the equation that is to be solved (the equivalent of $A \mathbf{s} = b$ in sparse regression), and providing the bias towards a bimodal strain distribution.

\begin{figure}
\includegraphics[width=1.0\linewidth]{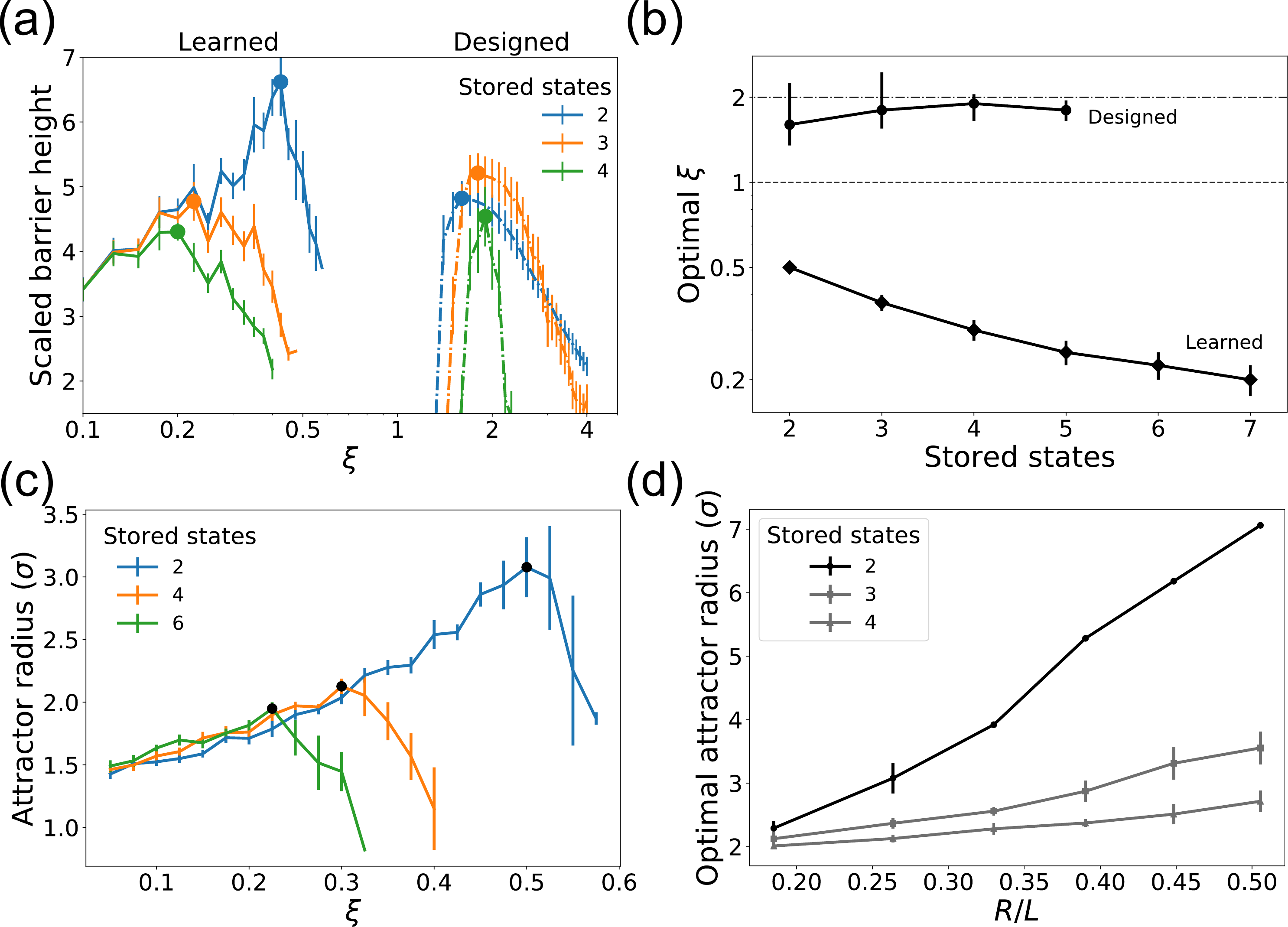}
\caption{Optimal non-linearity for learned and designed states. (a-b) 
Barrier heights around designed states are highest for $\xi^* \approx 2$ (linear springs) but highest for learned states at a specific non-linearity $0 < \xi^*<1$. Further, learning more states requires stronger non-linearity $\xi^*$. (c) We find similar results by quantifying learning quality by attractor size around stable states. 
(d) Learning rules that connect more distant nodes, i.e., larger range $R$ for $f(r)$ in Eq.~\ref{eq:LearningRule}, lead to larger attractor basins (see SI for details). $L$ is the system length.
\label{fig:AttractorQuality}}
\end{figure}

To test this analogy in larger elastic networks, we let a $N=100$ particle network learn two distinct states, and measured the strain in each spring after relaxing to one of the states (Fig.~\ref{fig:sparse}d). For non-linear springs $\xi<1$, we find a bimodal strain distribution - half of the springs are considerably strained, while the other half are at (approximately) zero strain. This result is in stark contrast to the designed minima with linear springs $\xi=2$, for which all springs are strained.

\subsection*{Optimal non-linearity}

The quality of both learning and design can be quantified by the attractor size and barrier heights around stored states. Large attractors and high energy barriers allow robust retrieval of states from a larger range of initial conditions.
These measures have long been used to quantify quality of learning in neural networks \cite{hertz1991introduction, amit1985storing, amit1985spin}.

We find that quality of designed and learned states, as measured by barrier heights, is highest at distinct $\xi^*$; see  Fig.~\ref{fig:AttractorQuality}a,b. The quality of designed states, for our simple design algorithm, is optimal for linear springs $\xi^* \approx 2$ and is relatively insensitive to the number of designed states. However, the optimal $\xi^*$ for learned states is $0 < \xi^* <1$ and varies with the number of learned states. We find similar results by measuring attractor radius instead of barrier heights (Fig.~\ref{fig:AttractorQuality}c). See Supplementary Note 3. 

Much as in sparse regression \cite{friedman2012fast,majumdar2010non}, the optimal $\xi^*$ for learning can be understood as a balance of two factors -- sparsity (smaller $\xi$) and convexity (larger $\xi$). Smaller $\xi$ leads to more sharply cusped energy contours in Fig.~\ref{fig:sparse}c and thus a stronger bias towards bimodal strain distributions with zero strain in some springs (i.e,. sparsity). However, smaller $\xi \to 0$ also leads to vanishing restoring forces outside the immediate vicinity of the unstrained configuration, creating a `golf course' landscape with vanishing attractors. 

Thus while smaller $\xi$ locally stabilizes each desired minimum using bimodal strain distributions, larger $\xi$ enlarges the attractor basin, making these minima easier to find. Similar considerations in canonical sparse regression problems select $\xi^* = 1$ as an optimal choice \cite{Lasso}.

The radius of spring connection $R$ plays an important role in setting the optimal $\xi^*$ value. We observe that the additional stabilizing contributions of the springs afforded at larger $R$ facilitates the optimal stabilization of the system at higher $\xi^*$, and thus with attractors of larger size, as seen in Fig.~\ref{fig:AttractorQuality}d (for more information see Supplementary Note 4). $L$ is the length scale of the system.

\begin{figure}
\includegraphics[width=0.95\linewidth]{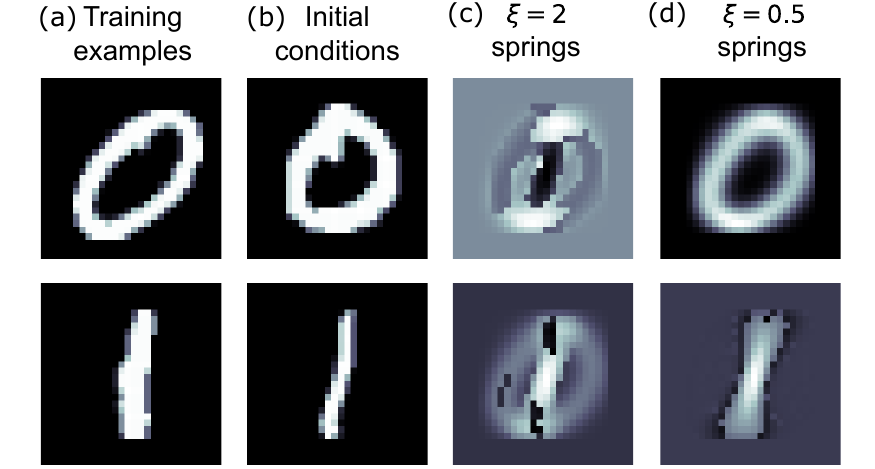}
\caption{Elastic networks learn to recognize handwritten digits. (a) 
Images representing two particle configurations that we wish to stabilize (adapted from MNIST). The $400$ pixel gray-scale values in each image are interpreted as positions of $400$ particles in 1-dimension. We learned a non-linear spring network using $5000$ randomly drawn examples of $0$s and $1$s each. (b) Learned networks are then tested by initializing at configurations corresponding to new unseen examples of `0' and `1'. (c) Linear networks fail to learn stereotyped states; initializing at each test example results in an unique uninterpretable state. (d) In contrast, non-linear networks learn two stereotyped states corresponding to `0' and `1' that are reliably retrieved in response to unseen examples of `0' and `1' from the MNIST database.
\label{fig:MNIST}}
\end{figure}

\subsection*{Pattern Recognition}
Finally, we ask whether our learned network with large robust attractors around the learned states can perform pattern recognition. To do this, we turn to the MNIST handwritten digits database \cite{lecun2010mnist}, and try to teach an elastic network to recognize the digits `0' and `1' from examples of these digits.

We trained the elastic network with $5000$ samples of the digits 0 and 1 each from the MNIST database in the following way; each $400$ pixel image was interpreted as a $1$-d configuration of $400$ particles by interpreting each pixel's gray-scale value as a particle’s position in the interval $[0,1]$. The particles in such a state are connected by elastic rods according to the learning rule in Eq.~\ref{eq:LearningRule}.
For $\xi< 1$, we find that the training generally creates two distinct large attractors corresponding to an idealized 0 and 1 respectively (Fig.~\ref{fig:MNIST}d).

We then test the network by using novel unseen examples of 0s and 1s from MNIST as initial conditions for the particles.  While these test images are not identical to any particular 0 or 1 used in training, the elastic network still retrieves the correct stored 0 or 1 state. Thus the non-linear $\xi \leq 1$ elastic network learns states 0 and 1 with sufficiently large attractors to accommodate the typical handwriting variations seen in the MNIST database.  

\section*{Discussion}

In this work we contrasted a design and a learning framework for creating multi-stable elastic networks. We found that continually learning novel states without overwriting existing states requires a specific non-linear elasticity $\xi \leq 1$. The learning model here relies on spontaneous growth of stabilizing rods between nearby nodes, a behavior displayed by microtubules \cite{Hess2017-gi}, DNA nanotubes \cite{mohammed2013directing} and other such seeded self-assembling tubes \cite{Li2005-lt, Hartgerink1996-uc, Blau2004-vy}.

The non-linearity $\xi$ plays a unique role as a material design parameter. Most material parameters (e.g., $l_{ij}$, $k_{ij}$ of springs here) encode information about desired states. But $\xi$ encodes an assumption about how information about desired states is distributed among parameters $l_{ij}$, $k_{ij}$ of different springs. Learning localizes information about each state to a subset of all springs. Hence stabilizing learned states requires $\xi < 1$, establishing states in which some springs are fully relaxed even if others are highly strained, i.e., the strain profile is sparse. In this way, the non-linearity $\xi$ is mathematically analogous to Bayesian priors in statistical regression that encode assumptions about the sparse nature of solutions. However, the elastic network here goes beyond the classic sparsity problem (Eq.~\ref{eq:LASSO}); the network has $2$-d spatial geometry absent in Eq.~\ref{eq:LASSO} and is more closely related to (unsolved) sparse reconstruction of $2$-d maps from pairwise distances between cities \cite{Montanari}. Consequently, we can explore how physical parameters with no analog in Eq.~\ref{eq:LASSO}, such as the maximum range of learned interactions $R$ (Fig.~\ref{fig:AttractorQuality}d) and spatial correlations between stored states, affect the optimal non-linearity $\xi$  (Supplementary Note 4).

Learning and design have complementary strengths, as seen before in neural networks and spin glasses. For example, Hopfield \cite{Hopfield} introduced neural networks that can learn arbitrary novel memories in sequence using a biologically plausible `Hebbian' learning rule. Gardner \cite{Gardner} showed that the same model has a higher memory capacity if we assume an optimally designed network in lieu of learning. However, Gardner's network can be designed only when all desired memories are known --- and must be redesigned from scratch to include new memories.

Similarly, in materials, design might be sufficient if all desired states are known beforehand and unlimited computational power is available, since design allows optimization over all design parameters. In contrast, learning is a physically constrained exploration of the same design parameters. However, such constrained exploration can be superior when the desired behaviors are not known \textit{a priori} and revealed only during use of the material itself. We hope the simple mechanical model studied here will stimulate further work on realistic learning rules that allow materials to acquire new functionalities on the fly.

\section*{Acknowledgments}
We thank Miranda Cerfon-Holmes, Sidney Nagel, Lenka Zdeborova for insightful discussions and Vincenzo Vitelli for a careful reading of the manuscript. We acknowledge NSF-MRSEC 1420709 for funding and the University of Chicago Research Computing Center for computing resources.

\bibliographystyle{unsrt}
\bibliography{LearningMemories_Citations,Paperpile}

\end{document}


\section*{Supplementary Note 1 - Design of multiple stable states with linear and non-linear springs}

As a simple model for weakly strained elastic materials, linear (Hookean) springs are often used for theoretical constructions of elastic networks. Each of the two nodes connected by a linear spring of stiffness $k$ and rest length $l$, and separated by distance $r$, feels a force $\vert \textbf{F} \vert = k \vert r-l \vert$. The energy associated with the straining of the spring is $E=\frac{1}{2}k(r-l)^2$.

Suppose we construct a network with $N$ nodes embedded in $d$-dimensional space. Each 2 nodes (located at $\textbf{x}_i, \textbf{x}_j$) are connected by a linear spring of stiffness $k_{ij}$ and rest length $l_{ij}$. The energy of the elastic network is

\begin{align}
E(\{\textbf{x}\}) = \frac{1}{2} \sum_{i=1}^{N} \sum_{j=i+1}^{N} k_{ij}(r_{ij}-l_{ij})^2,
\label{eq:SI-LinEnergy}
\end{align}

where $r_{ij} \equiv \vert \vert \textbf{x}_i - \textbf{x}_j \vert \vert $ are the distances between nodes. The stable configurations (minima) of this energy function are found by equating the gradient of Eq.~\ref{eq:SI-LinEnergy} with respect to node positions to zero:

\begin{align}
0 = \partial_{\textbf{x}_a} E = \sum_{i=1}^{N} \sum_{j=i+1}^{N} k_{ij}(r_{ij}-l_{ij}) \frac{\partial r_{ij}}{\partial\textbf{x}_a}.
\label{eq:SI-LinGradient}
\end{align}

\begin{figure*}	
\includegraphics[width=0.5\linewidth]{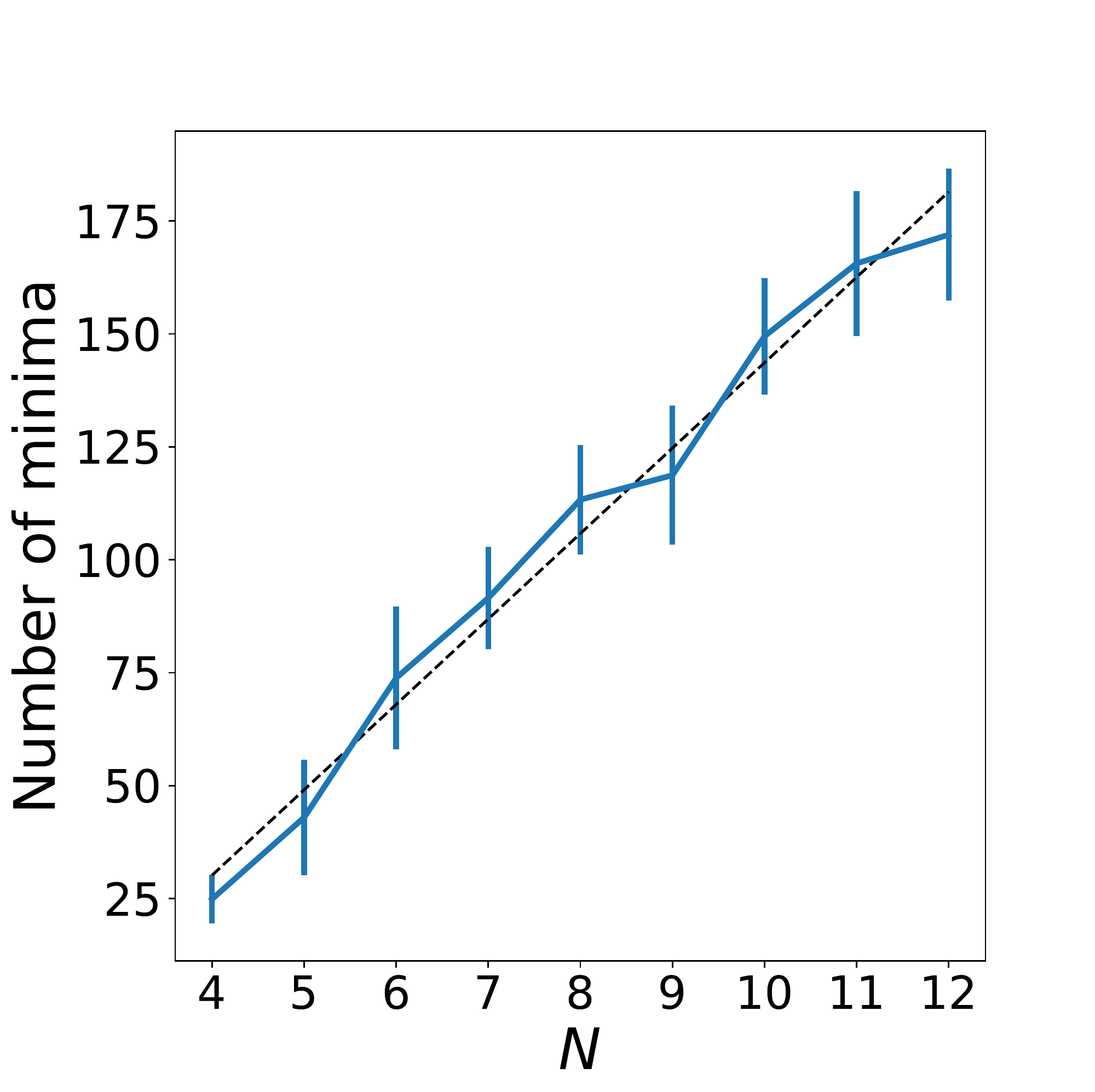}
\caption{Number of stable configurations in a network of linear springs grows linearly with the size of the system.
\label{fig:NumMinima}}
\end{figure*}

This procedure gives $N d$ equations that have to be satisfied simultaneously for the $N d$ node coordinates. Note that Eq.~\ref{eq:SI-LinGradient} is not linear in node coordinates, as the distances in dimension $d$ are computed by $r_{ij}=\sqrt{\sum_d (x_{i,d}-x_{j,d})^2}$ (manifestly nonlinear in $x_i$ for $d>1$, but even for $d=1 \rightarrow r_{ij}=\vert x_i-x_j\vert$). Due to the nonlinear relation of $r_{ij}$ to $x_i,x_j$, multiple solutions $\{\textbf{x}^\star\}$ can satisfy Eq.~\ref{eq:SI-LinGradient} simultaneously. Even though one still needs to check the second derivative at the proposed configuration $\{\textbf{x}^\star\}$ to test if it is a stable minimum, in practice we find that there indeed exist multiple stable points for two-dimensional embeddings. Simulating small systems with up to $12$ nodes in $2d$, we find that the number of minima scales linearly with node number (Fig.~S\ref{fig:NumMinima}).


These multiple minima in the potential energy landscape, if moved around, could be utilized to program the desired stable configurations. This is possible by careful choice of the stiffness values $k_{ij}$ and rest lengths $l_{ij}$ of all springs. Note that even tough Eq.~\ref{eq:SI-LinGradient} is nonlinear in node positions $\{\textbf{x}\}$, it is linear in both $k_{ij}$ and $a_{ij}\equiv k_{ij}\cdot l_{ij}$. Suppose we want to solve the system of equations \ref{eq:SI-LinGradient} for $M$ different node configurations denoted by $\{\textbf{x}\}^m$, giving rise to distance matrices $r_{ij}^m$. Solution to such linear systems of equations can generally be found if the number of equations ($Nd M$) is less than the number of variables ($0.5N^2d$). To design linear springs with multiple desired stable points, we thus numerically solve Eq.~\ref{eq:SI-LinGradient} simultaneously for the desired configurations $\{\textbf{x}\}^m$ to get the values of $k_{ij},l_{ij}$, and then check that the obtained elastic network is indeed stable in all of these configurations.

The particular algorithm discussed above is only defined for linear springs with $\xi=2$, as defined in the main text. Still, a design protocol for spring-node systems with any value of non-linearity $\xi$ is possible. With non-linear springs the force balance of Eq.~\ref{eq:SI-LinGradient} becomes:

\begin{align}
0 = \partial_{\textbf{x}_a} E \sim \sum_{i=1}^{N} \sum_{j=i+1}^{N} k_{ij}(r_{ij}-l_{ij})^{\xi-1} \frac{\partial r_{ij}}{\partial\textbf{x}_a},
\label{eq:SI-NonLinGradient}
\end{align}

which is unfortunately non-linear in the rest lengths $l_{ij}$. In similar spirit to the above algorithm, we minimize the sum-squared of all $NdM$ equations due to the set of Eq.~\ref{eq:SI-NonLinGradient} over the design parameters $k_{ij}, l_{ij}$. If minimization succeeds in finding perfect (zero) solutions, it gives sets $k_{ij}, l_{ij}$ for which the nodes feel very little force in all of the $M$ stable states. We can then numerically check whether these states are stable.

The capacity of designed networks to store multiple stable states $M_C$ is expected to scale linearly with system size (number of nodes $N$). This idea arises as stabilizing $M$ states requires the simultaneous satisfaction of $NdM$ constraints using $0.5N^2d$ parameters as discussed above.  These two numbers match for a critical number of states $M_C\sim N$, and for $M>M_C$ no solution exists in general. Unfortunately, this prediction is difficult to corroborate numerically due to the computationally NP-hard nature of the design problem.

\section*{Supplementary Note 2 - Energy model for nonlinear springs}

The main text establishes that to enable the learning paradigm to store multiple stable states in an elastic networks, one needs to utilize nonlinear springs with certain properties. Most importantly, if a spring is to hold information about one configuration associated to it, the spring should apply a strong force only when the system is close to its associated configuration. One simple way to parametrize such forcing is to use a spring whose force when pulled away from the preferred length is $F\sim(r-l)^{\xi-1}$. Clearly, if one chooses $\xi=2$, the limit of linear springs is obtained once more, where the force gets stronger the further the spring is strained.

If one chooses $0<\xi<1$, the spring's response weakens as it is strained. Unfortunately such springs are nonphysically singular for $r=l$. One way to counter this singularity is to introduce a linear "core" spring, with some length scale $\sigma$, such that the spring behaves like a linear spring for $\vert r-l \vert <\sigma$, and non-linearly otherwise. If we define a non-dimensional strain $u\equiv (r-l)\sigma^{-1}$, the energy of such a spring can can be written as:

\begin{align}
E(u) = \frac{1}{2} k \sigma^\xi \cdot\frac{u^2}{(1+u^2)^{1-0.5 \xi}} ,
\label{eq:SI-NonLinearSpring}
\end{align}

with $r$ the spring length, $k$ stiffness, $l,\sigma$ the rest length and "core" size, respectively. The prefactor $\sigma^\xi$ is chosen so that the long range forces $u\rightarrow\infty$ are independent of the core size $\sigma$, and that the $\xi=2$ limit is the desired linear spring. In this model, spring non-linearity is controlled by the exponent $\xi$, defined in a way to recapitulate the behavior of regularizers in optimization problems. A choice of $\xi = 2$ gives rise to linear springs, akin to ridge regularization, while $\xi = 1$ gives long range constant forces $E\sim u$, similar to LASSO regularization. The extreme limit $\xi=0$ defines springs whose energy is a Lorentzian. Outside the core region, such springs exert forces that diminish quickly as $F\sim u^{-1}$. In general, the force due to the nonlinear springs is

\begin{align}
F(u) = k \sigma^{\xi-1} u \cdot\frac{1+0.5\xi u^2}{(1+u^2)^{2-0.5 \xi}}.
\label{eq:SI-NonLinearSpringForce}
\end{align}

\begin{figure*}	
\includegraphics[width=0.6\linewidth]{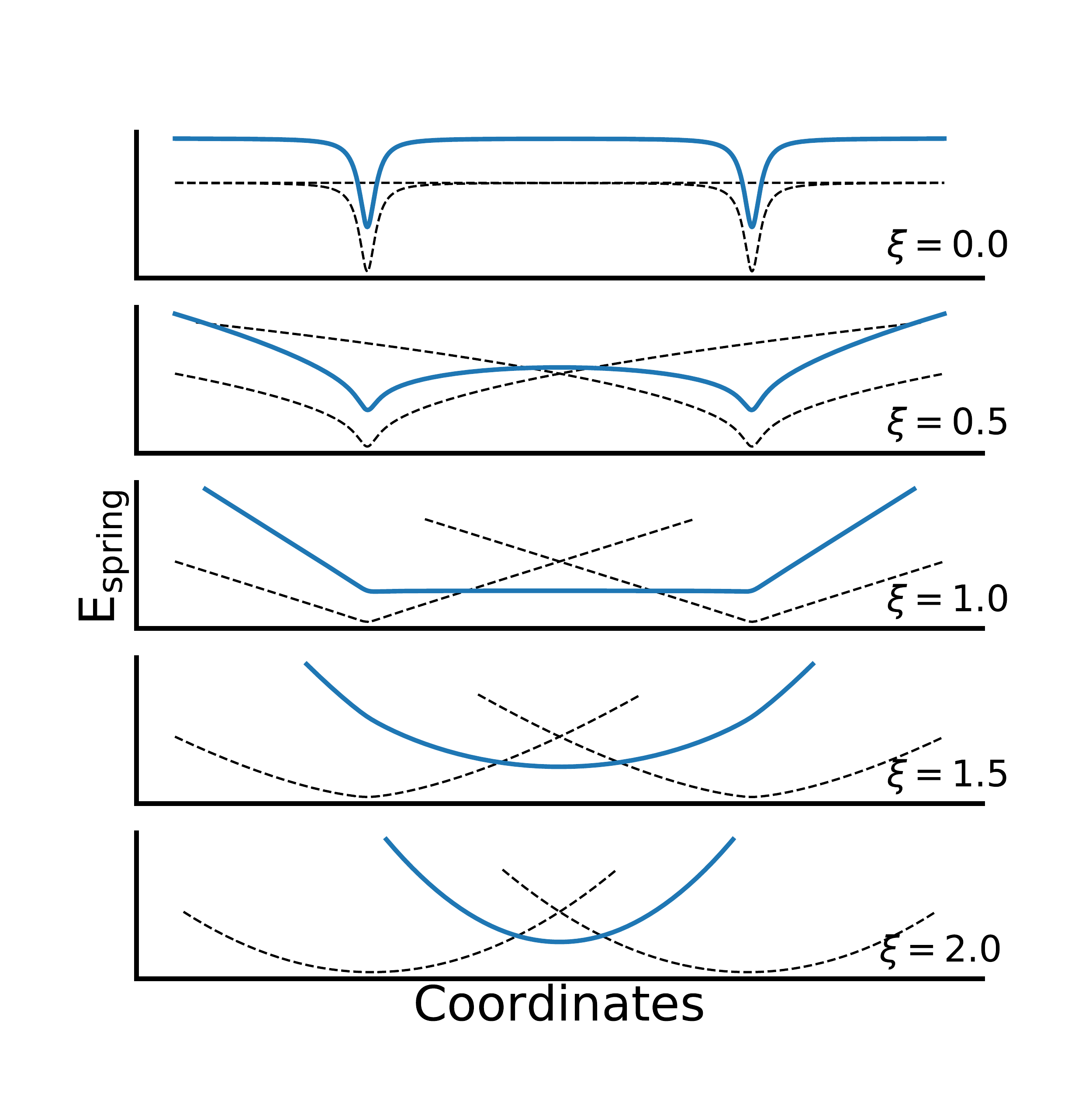}
\caption{The sum energy of two springs goes through a transition at $\xi=1$. The energy minimum of each spring is preserved for $\xi<1$, while these minima are overwritten for $\xi>1$. In essence, the information on minima of $\xi<1$ springs is stored with each individual spring. (Black dotted lines correspond to the individual potentials of two nonlinear springs with given $\xi$, bold blue lines show the sum of the two potentials, shifted up for clarity).
\label{fig:SI-XiTransition}}
\end{figure*}

The crucial property of this family of spring potentials is the force behavior at large strains, far beyond the core $u \gg 1$. At large strains the force applied by the springs is $F\sim u^{\xi-1}$, a form which goes through an important transition at $\xi = 1$. For springs with $\xi > 1$, the restoring force grows with strain, while for $\xi < 1$ the force diminishes. This transition causes an important change of behavior when such spring potentials are summed together, as shown in Fig.~S\ref{fig:SI-XiTransition}. The minima of individual springs are preserved for $\xi<1$, while these minima are overwritten for springs with $\xi>1$. We conclude that only springs with $\xi < 1$ (or more generally, springs whose force diminishes with range) enable the learning paradigm described in the main text. In learning, we would like the information about each stored state to be localized to a subset of springs, and that adding more springs for new states does not overwrite the previously stored information. Figure~S\ref{fig:SI-XiTransition} clarifies that springs whose force grow with strain are completely unfit for this purpose.

\newpage
\section*{Supplementary Note 3 - Numerical exploration of mechanical networks}

Testing predictions about learning elastic networks requires the numerical construction of such networks, and the ability to explore their potential energy landscape. This section describes some of the technical aspects involved in simulating these networks and deducing their properties. The codes to produce and study the elastic networks is implemented in Python and available upon request.

\subsection*{Network construction}

The elastic networks simulated for this work consist of $N$ nodes embedded in a $2d$ box of size $1\times 1$. For each desired system configuration (stored state), node positions are sampled uniformly at random within the boundary of the box. Each multi-stable system of this type with $M$ states is thus described by $M\times N \times 2$ positions in the range $[0,1]$. Springs are attached between pairs of nodes according the paradigm studied (design, learning). 

For the study of design, we fully connect all pairs of nodes in the system with linear springs ($\xi=2$). These springs are chosen to take into account all of the desired states simultaneously. The choice of springs (stiffness and rest length values) is made by solving the set of equations \ref{eq:SI-LinGradient} in Supplementary Note 1. Construction of fully connected designed networks with non-linear springs ($\xi\ne 2$) is facilitated by optimizing forces at the desired stored states (Supplementary Note 1).

System with learned states are constructed by attaching a set of springs between pairs of nodes for each stored state. We generally do not fully connect the nodes, instead opting to connect a spring between nodes within a certain chosen distance $R$, as outlined in the main text. All springs in this paradigm have the same spring stiffness $k$, core size $\sigma$ and non-linearity parameter $\xi$. The springs only differ in their rest length, chosen so that the springs are relaxed in their respective state. Thus, learning is 'easy' in the sense that no computation is required to choose the new set of springs in new stored states. This suggests learning can be performed by a rather simple, physically passive system, whose time evolution depends only on its current configuration.

\subsection*{Estimation of attractor size and barrier height}

When $M>1$ states are encoded into a network, it is of immediate interest to check whether these states are stable at all. We define a stable state $\vec{X}^{(m)}$ ($N\times 2$ spatial vector) by the following requirement: when the system is released from $\vec{X}^{(m)}$ and allowed to relax to a nearby stable minimum of the potential energy landscape, the relaxed configuration $\vec{X}_*^{(m)}$ is close in configuration space to $\vec{X}^{(m)}$. We consider states to be preserved if the average displacement per degree of freedom after relaxation is much smaller than the size of the box. The potential core size $\sigma$ is used as this stability cutoff $\frac{\vert\vert \vec{X}_*^{(m)} - \vec{X}^{(m)} \vert\vert}{2N} < \sigma$, where the typical core size is $\sim 1\%$ of the box size. If the different encoded states pass this test, we say that the states are stable, and the encoding was successful. See Supplementary Note 4 for more details on the stability of stored states.

In an effort to find optimal schemes for storing stable states in elastic networks, basic stability does not suffice, and we require additional measures of merit. A natural approach is to study the attractor basins of the encoded states, specifically their spatial extent and the energetic barriers surrounding them. The larger the attractor basin, the configuration can more reliably be retrieved when the system is released farther away from its minimum. High energy barriers surrounding the state basins improve their stability when the system is subjected to finite temperatures or other types of noise.

Unfortunately both attractor size and energy barrier are non-local properties of the attractor, requiring many high-dimensional measurements away from the stable state. Rather than exhaustively studying the attractor basin shape and height, we employ a procedure as follows: at the stable state, choose a random direction and take the system a small amount in that direction. Relax the system from the new position and verify whether it relaxed into the same stable state. If so repeat the last step, but take the system slightly farther away in the same direction as before. Repeat these steps until the system no longer relaxes to the initial state, but instead reaches another stable point of the landscape. Measuring the distance required to move the system in order to escape the attractor, and the energy at that distance, furnishes an estimate of both the attractor size and the energy barriers around it. We repeat the above process to average the results over many different random directions in configuration space.

An important correction is needed for the above estimation, in particular for the flatter spring potentials $\xi \ll 1$. Attractors arising from these potentials tend to be very flat far from the core region $\sigma$ surrounding each stored state. Although flat regions mathematically belong to some attractor basin, releasing the system in these regions will require long relaxation times, and relaxation dynamics are highly unstable to external noise. We therefore define a 'useful' attractor, such that the gradient that leads relaxation towards the stable point is large enough. In practice, we cut-off the attractor defined by the previous algorithm when the relaxation force is smaller than a fraction ($\sim 0.5$) of the typical force within the core distance $\sigma$. The inclusion of this force (gradient) requirement gives rise to an optimal non-linearity value $0<\xi<1$ for learned states, as shown in the main text.

\section*{Supplementary Note 4 - Stability of learned states}

In the main text we established the usefulness of learning with non-linear springs as a means of programming multiple stable states into an elastic network. In this section we discuss some limitations of this idea, such as the finite capacity of node-spring networks, and the effect of connectivity within a state and correlations between states on the quality of learning. 

\subsubsection*{Storing capacity}

Nonlinear spring networks (with $\xi<1$) can stabilize multiple states through sparsity - springs associated with a certain state dominate the response of the network when it is situated close to that state. Springs associated with other states are highly stretched, yet apply small forces that further diminish at high strains. Still, force contributions of springs unrelated to the desired state are finite and act to destabilize that state. 

\begin{figure*}	
\includegraphics[width=0.95\linewidth]{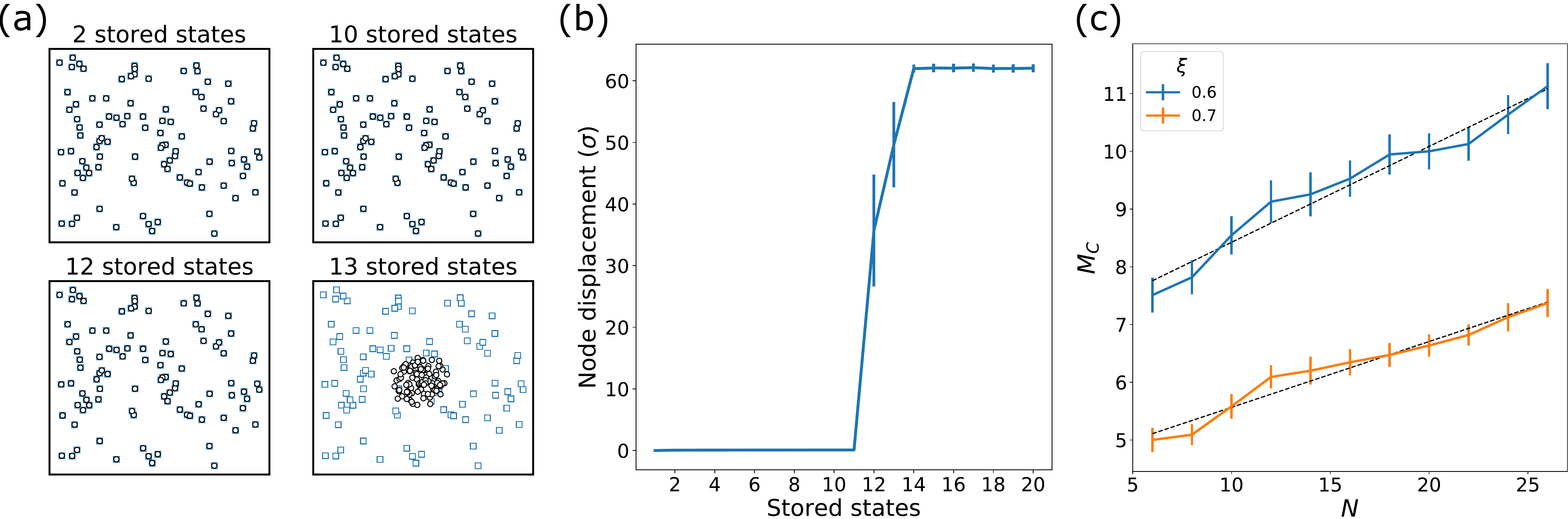}
\caption{Programming stored states using the learning paradigm exhibits finite capacity. a) Each stored state is affected by springs associated with other states. Initially the new springs have a small effect and the state remains a stable attractor. However, eventually states destabilize due to the forces exerted by the other stored states (Blue squares denote a certain stored state, black circles show the nearby stable configuration). This state fails when 13 statess are simultaneously encoded ($N=100,\ \xi=0.6$). b) When node displacement is averaged over stored states, we observe a sharp failure of all stored states close to a specific load, defined as the \textit{capacity} ($12-13$ states in this case). c) Capacity scales linearly with system size $N$.
\label{fig:SI-Cap}}
\end{figure*}

The learned networks studied in this work exhibit destabilization of learned states due to the effect of springs associated with other stored states. Figure~S\ref{fig:SI-Cap}a shows a typical scenario observed in these networks, where a desired state is stable when the overall number of learned states is low. Then, an abrupt threshold (capacity) is passed after which the state destabilizes completely and the system relaxes into a configuration that looks completely different from the desired stored state. Generically, all learned states fail in this way at a similar capacity value (Fig.~S\ref{fig:SI-Cap}b). This capacity is well-defined and observed to depend on the parameters of the system, such as size $N$ and non-linearity $\xi$.

Let us now argue for a scaling relation of the storing capacity. Suppose a system of $N$ nodes is used to learn $M+1$ states. In configurations close to state $1$, $N$ springs will apply a stabilizing force $F_S$, while the rest $N\times M$ springs will act to destabilize the state with force $F_{DS}$. All stabilizing springs provide a force in the same stabilizing direction such that $F_S\sim N$. If we assume the $N\times M$ destabilizing forces due to unrelated springs are randomly oriented and similar in magnitude, the total destabilizing force would behave like a random walk and have a magnitude $F_{DS}\sim\sqrt{N\times M}$. The capacity of the system is reached when the magnitude of the destabilizing force is equal to that of the stabilizing force, so that

\begin{align}
F_{DS}(M_C)\sim F_S \rightarrow M_C\sim N.
\label{eq:SI-NScaling}
\end{align}

The capacity of a learning network is expected to scale linearly with system size, similarly to other Hopfield-inspired learning models \cite{Hopfield}. This prediction was tested in networks with of sizes $N=6-26$ and for several values of the non-linearity $\xi$. Results shown in Fig.~S\ref{fig:SI-Cap}c are consistent with the linear scaling suggested above. Theoretical arguments of a similar nature suggest another scaling relation $M_C\sim \exp(-\xi)$, also in agreement with numerical data. However, we regard the capacity dependence on non-linearity to be of lesser interest, as other metrics for quality of encoding (barrier height and attractor size), discussed in the main text, are more important for the robustness of learning.

\subsubsection*{Connectivity of nodes}

\begin{figure*}	
\includegraphics[width=0.9\linewidth]{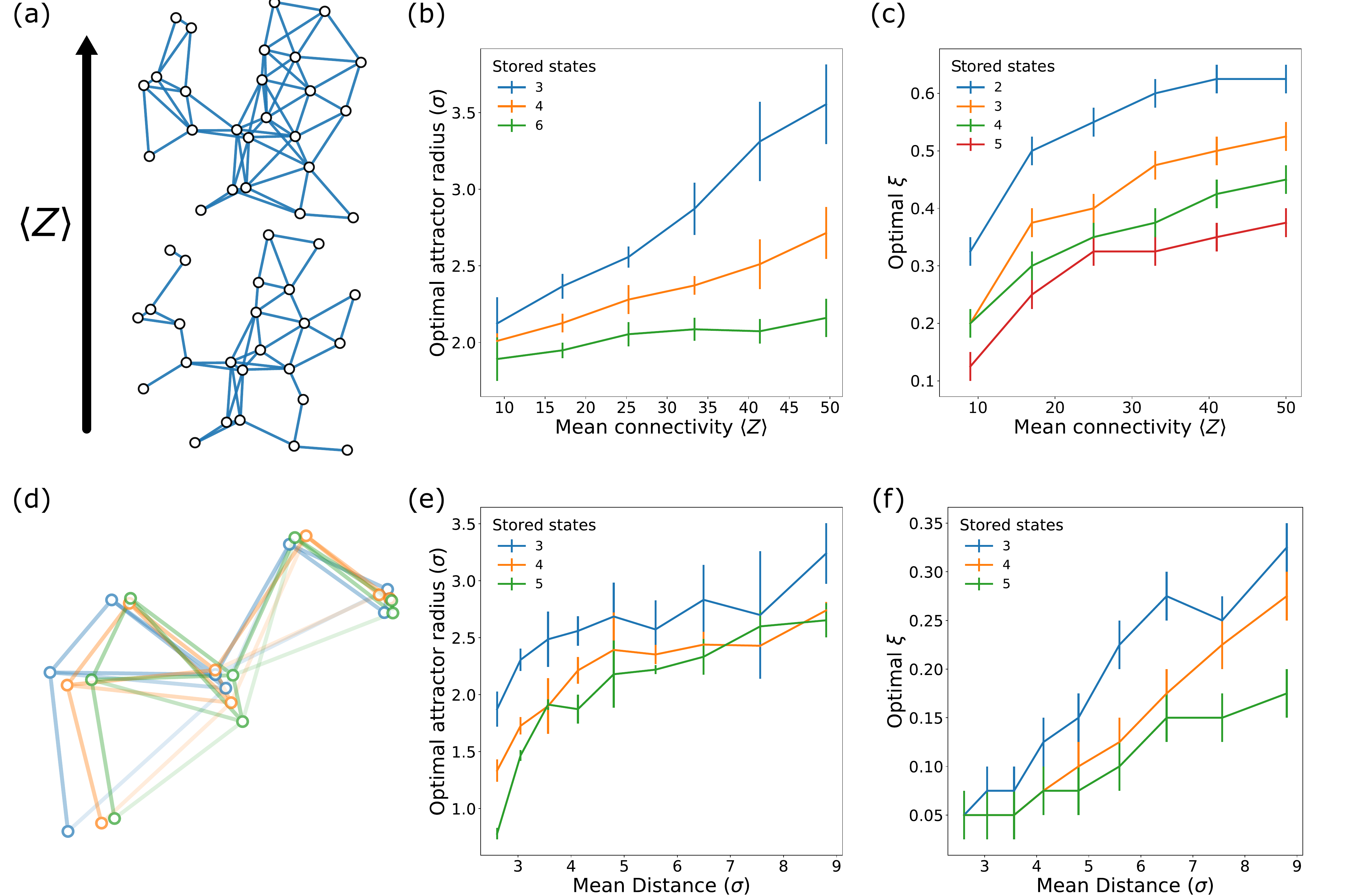}
\caption{Effects of node connectivity and state similarity on the quality of encoding the stored states. (a-c) Connectivity between nodes $\langle Z \rangle$ increases with the effective connection radius $R$. We find that the more internally connected a state is, the larger its attractor size, and higher the optimal value of the non-linearity parameter $\xi$ ($N=100$). (d-f) Trying to store similar states is more difficult than random states. When the mean distance between nodes in successive states are small, attractors basin are also small, and successful encoding requires small values of $\xi$, and thus flat potentials ($N=100$).
\label{fig:SI-CS}}
\end{figure*}

It is well known that the rigidity of elastic networks strongly depends on node coordination - the number of springs connected to the different nodes. Rigid networks are characterized by coordination numbers exceeding the Maxwell condition \cite{maxwell1864calculation}. Then, a stable state of the over-constrained network can be understood as a minimum point of the energy landscape constructed of the spring potentials. Further increasing the coordination of nodes - or their connectivity to other nodes, usually results in stable states surrounded by higher energy barriers. 

This argument suggests an intriguing possibility, that increasing connectivity in learned network may improve the stability and quality of the state storage. Such an outcome is possible as the act of adding more non-linear ($\xi<1$) springs associated with a certain stored state is not expected to significantly alter the state itself, since the rest lengths are chosen to stabilize this state. On the other hand, the extra springs may increase the height of energy barriers surrounding the state, making it more stable against temperature and noise. Furthermore, increasing connectivity may also enlarge the attractor regions of stored states, as the extra constraints induced by the new springs may suppress 'distractor' states (spurious energy minima due to partial satisfaction of the frustrated interactions). 

In the context our learning paradigm, connectivity is controlled by the effective radius of rod growth $R$ defined in the main text. If states are constructed by randomly placing $N$ nodes in a $d$-dimensional square box of length $L$, it is easy to see that the average connectivity scales as $\langle Z \rangle \sim NR^d$ while $R\ll L$. We use $N=100,\ \xi<1$ networks to test the effect of node connectivity on the attractor size of stored states. Results presented in Figure~S\ref{fig:SI-CS}(a-c) verify that the quality of state storage, as measured by the attractor size of states, improves with their connectivity.

\subsubsection*{State similarity}

In most of this work we considered stored states that are completely uncorrelated between themselves, i.e. the position of a node in each stored state is independent of its position in other states. In practice, it might be easier conceive of elastic networks whose different stable states are not too different from one another, in which neighboring nodes in one configuration will remain neighbors in other configurations. Furthermore, some applications (e.g. classification of similar objects) may require different stored states to be correlated to differing extents. In general, encoding correlated (i.e. similar) states is expected to negatively affect the stability of these states and their quality (as measured by attractor properties as size and barrier heights).

To test the impact of similarity between states, we embedded a $N=100$ network with states in which the average displacement of nodes in successive states was controlled. Figure~S\ref{fig:SI-CS}(d-f) shows that the larger the difference between states, the larger their respective attractor sizes. In addition, larger differences between states allows their stabilization at higher $\xi$ values, which is expected to improve the heights of energy barriers surrounding them and further suppress distractor states. Still, we show that it is possible to encode multiple states in elastic networks, even when the average difference between stored states is a small multiple of $\sigma$ (the potential core size, within which states are indistinguishable).

\bibliographystyle{unsrt}
\bibliography{LearningMemories_Citations}